\documentclass[twocolumn,aps,prd,superscriptaddress,showpacs,floatfix]{revtex4}

\usepackage{epsfig,bm,feynmf}

\usepackage{graphics}

\usepackage[normalem]{ulem} 
\renewcommand\sout{\bgroup \color{red} \ULdepth=-.5ex \ULset}
\usepackage[dvips]{color} 

\def\ltap{\ \raise.3ex\hbox{$<$\kern-.75em\lower1ex\hbox{$\sim$}}\ }
\def\gtap{\ \raise.3ex\hbox{$>$\kern-.75em\lower1ex\hbox{$\sim$}}\ }

\begin{document}



\title{Exotic nuclei with open heavy flavor mesons}

\author{Shigehiro Yasui}
\email{yasuis@post.kek.jp}
\affiliation{Institute of Particle and Nuclear Studies, High Energy Accelerator Research Organization (KEK), 1-1, Oho, Ibaraki, 305-0801, Japan}
\author{Kazutaka Sudoh}
\email{k-sudoh@nishogakusha-u.ac.jp}
\affiliation{Nishogakusha University,
         6-16, Sanbancho, Chiyoda, Tokyo, 102-8336, Japan}

\begin{abstract}
We propose stable exotic nuclei bound with $\bar{D}$ and $B$ mesons with respecting heavy quark symmetry.
We indicate that an approximate degeneracy of $\bar{D}$($B$) and $\bar{D}^{*}$($B^{*}$) mesons plays an important role, and discuss the stability of $\bar{D}N$ and $BN$ bound states.
We find the binding energies 1.4 MeV and 9.4 MeV for each state in the $J^{P}=1/2^{-}$ with $I=0$ channel.
We discuss also possible existence of exotic nuclei $\bar{D}NN$ and $BNN$.
\end{abstract}



\maketitle

Researches of exotic nuclei have been one of the most interesting subjects in nuclear physics.
Recently there have been much progress in studies of proton- and neutron-rich nuclei with large isospin \cite{Tanihata:1986kh}, and hypernuclei and kaonic nuclei with strangeness \cite{Akaishi:2002bg,Gal:2009zk}.
Exotic nuclei are useful to study various aspects of non-perturbative QCD, such as exotic hadrons, nuclear force, dense and hot matter, and so forth.
They also provide us fundamental information for astrophysics.

For enlargement of our knowledge of exotic nuclei, the variety of multi-flavor is now going to be extended to heavier flavors of charm and bottom.
So far, several studies have been advocated for exotic nuclei with charmed baryons \cite{Dover:1977jw} and charmed mesons \cite{Tsushima:1998ru}, in which the interaction is based on SU(4) flavor symmetry as a straightforward extension from strangeness to charm.
Recently, inspired by the successful application of the chiral dynamics with approximate SU(3) chiral symmetry to strangeness sector \cite{Gal:2009zk}, the extended version with SU(4) chiral symmetry has been applied to charm sector \cite{Lutz:2003jw}.
However, the dynamics would drastically change in the system with heavy quarks, since it realizes not only chiral symmetry but also a new symmetry, namely a heavy quark symmetry \cite{Manohar:2000dt}.
This symmetry has been successfully applied to heavy flavor hadrons \cite{Isgur:1991wq,Swanson:2006st,Voloshin:2007dx}, and other exotic hadrons \cite{AlFiky:2005jd,Manohar:1992nd}. 

In this work, we investigate exotic nuclei bound with an open heavy flavor meson, $D$ or $B$ meson, with respecting the heavy quark symmetry. 
This approach would provide us new knowledge of exotic nuclei.
Such exotic nuclei will be experimentally accessible at future high-energy hadron facilities such as J-PARC (Japan Proton Accelerator Research Complex) and GSI (Gesellschaft f\"ur Schwerionenforschung) \cite{Lutz:2009ff}. 

One of the remarkable features of the heavy quark symmetry is a degeneracy of pseudoscalar and vector mesons as seen in small mass splitting between $D$ and $D^{*}$ mesons, and $B$ and $B^{*}$ mesons.
Therefore, both pseudoscalar and vector mesons are considered as fundamental degrees of freedom in the dynamics.
The picture for heavy quark is completely different from the picture for strange quark, in which approximate chiral symmetry is realized.
In the strangeness sector, only $K$ mesons are important due to nearly massless Nambu-Goldstone bosons, and hence $K^{*}$ mesons are almost irrelevant.

The interaction between $D$ ($D^{*}$) mesons and nucleons is thus qualitatively different from that between $K$ mesons and nucleons.
In the charm sector, approximate degeneracy of $D$ and $D^{*}$ mesons provides both $\pi DD^{*}$ and $\pi D^{*}D^{*}$ couplings, and it induces one pion exchange potential (OPEP) in the $t$-channel of the $DN$-$D^{*}N$ and $D^{*}N$-$D^{*}N$ scatterings.
In the strangeness sector, the absence of $K^{*}$ mesons leads non-existence of OPEP.
Instead, the dominant force is provided by the Weinberg-Tomozawa type interaction.
Therefore, we consider that the $DN$ and $D^{*}N$ interaction is concerned to a long distance force with a scale of inverse of $\pi$ meson mass, while the $KN$ interaction is a short distance force with a scale of inverse of $\omega$ and $\rho$ meson masses.
Such a picture will be applied to $B$ and $B^{*}$ mesons with more accuracy.

In this Letter, we discuss the $\bar{D}N$ and $BN$ states as the smallest system for an open heavy flavor meson and nucleons.
The $s$- and $u$-channels in the $\bar{D}N$ and $BN$ scattering would be suppressed, because the former includes a pentaquark with a heavy quark $Q$, and the latter requires a creation of a $Q\bar{Q}$ pair.
Consequently, the OPEP in the $t$-channel dominates for this system.
In contrast, the $DN$ and $\bar{B}N$ states must be more complicated because of the quark annihilation process.
For these reasons, we concentrate on the $\bar{D}N$ and $BN$ states here.
As the smallest nuclei with $\bar{D}$ and $B$ mesons, we address possible $\bar{D}NN$ and $BNN$ states with baryon number two.

The heavy quark symmetry with light quark chiral symmetry
 provides the vertex of $\pi$ mesons and open heavy flavor mesons $P$ ($D$ or $B$) and $P^{*}$ ($D^{*}$ or $B^{*}$)
\begin{eqnarray}
{\cal L}_{\pi H H} = g \, \mbox{tr} \bar{H}_{a}H_{b}\gamma_{\nu}\gamma_{5}A_{ba}^{\nu},
\label{eq:piHH}
\end{eqnarray}
where the multiplet  field $H$ of $P$ and $P^{*}$ is defined by
\begin{eqnarray}
H_{a} = \frac{1+/\hspace{-0.5em}v}{2} \left[ P_{a\mu}^{*} \gamma^{\mu} - P_{a} \gamma_{5} \right],
\end{eqnarray}
with the velocity $v$ of the mesons \cite{Isgur:1991wq}.
The conjugate field is $\bar{H}_{a} = \gamma_{0} H_{a}^{\dag} \gamma_{0}$, and the index $a$ denotes up and down flavors.
The axial current is given by
$
A_{\mu} \simeq \frac{i}{f_{\pi}} \partial_{\mu} {\cal M}
$
with 
\begin{eqnarray}
{\cal M} = \left(
\begin{array}{cc}
 \frac{\pi^{0}}{\sqrt{2}} & \pi^{+} \\
 \pi^{-} & -\frac{\pi^{0}}{\sqrt{2}}
\end{array}
\right),
\end{eqnarray}
where $f_{\pi}=135$ MeV is the pion decay constant.
The coupling constant $|g|=0.59$ for $\pi P P^{*}$ is determined from the observed decay width  $\Gamma = 96$ keV for $D^{*} \rightarrow D\pi$ \cite{Amsler:2008zzb}. 
The coupling of $\pi P^{*}P^{*}$, which is difficult to access from experiments, is automatically determined thanks to the heavy quark symmetry.
Note that the coupling of $\pi P P$ does not exist due to the parity conservation.
The coupling constant $g$ for $\pi B B^{*}$ would be different from one for $\pi D D^{*}$ because of $1/m_{Q}$ corrections with the heavy quark mass $m_{Q}$ \cite{Cheng:1993gc}.
The recent lattice simulation in the heavy quark limit suggests the closed value adopted above \cite{Ohki:2008py}.
This would allow us to use the common value for $D$ and $B$.

In order to derive the OPEP between $P^{(*)}$ and $N$ in non-relativistic form, we set $v=(1,\vec{0})$ in Eq.~(\ref{eq:piHH}), and obtain the $\pi PP^{*}$ and $\pi P^{*}P^{*}$ vertexes
\begin{eqnarray}
{\cal L}_{\pi P P^*}  \!&=&\! \sqrt{2} \, \frac{g}{f_{\pi}}\,
\left[ (\vec{\varepsilon}\!\cdot\!\vec{\nabla}) \, (\vec{\pi}\!\cdot\!\vec{\tau}) + (\vec{\varepsilon}^{\hspace{0.2em}*}\!\cdot\!\vec{\nabla}) \, (\vec{\pi}\!\cdot\!\vec{\tau}) \right], \label{eq:piPP*} \\
{\cal L}_{\pi P^* P^*} \!&=&\! -\sqrt{2} \, \frac{g}{f_{\pi}} \, (\vec{T} \!\cdot\! \vec{\nabla}) (\vec{\pi}\!\cdot\!\vec{\tau}), \label{eq:piP*P*}
\end{eqnarray}
where the polarization vector of $P^{*}$ are defined by
$\vec{\varepsilon}^{\hspace{0.2em}(\pm)} \!=\! \left(\mp 1/\sqrt{2}, \pm i/\sqrt{2}, 0 \right)$ and
$\vec{\varepsilon}^{\hspace{0.2em}(0)} \!=\! \left(0, 0, 1\right)$,
$\vec{\pi}$ is the pion field,
and the spin-one operator $\vec{T}$ is defined by $T_{\lambda' \lambda}^{i}=i \varepsilon^{ijk} \varepsilon_{j}^{(\lambda')\dag} \varepsilon_{k}^{(\lambda)}$.
The $\pi NN$ vertex is
\begin{eqnarray}
{\cal L}_{\pi NN} = - \frac{g_{\pi NN}}{m_{N}} \chi_{s'}^{\dag} \, \vec{\sigma} \!\cdot\! \vec{\nabla} \frac{\vec{\pi} \!\cdot\! \vec{\tau}}{2} \, \chi_{s},
\label{eq:piNN}
\end{eqnarray}
where $\chi_{s}$ is the nucleon field with spin $s$, $m_N$
 is the nucleon mass, and $g_{\pi NN}^{2}/4\pi=13.5$ is the coupling constant.
From the vertexes (\ref{eq:piPP*}), (\ref{eq:piP*P*}), and (\ref{eq:piNN}), the OPEPs in the $PN$-$P^{*}N$ and $P^{*}N$-$P^{*}N$ scatterings are given by
\begin{eqnarray}
V_{PN \rightarrow P^{*}N} &\!=\!& - \frac{g g_{\pi NN}}{\sqrt{2}m_{N}f_{\pi}} \frac{1}{3}  \label{eq:pot_PNP*N_ff} \\
&&\hspace{-0em} \times \left[ \vec{\varepsilon}^{\,(\lambda)\dag}\!\cdot\!\vec{\sigma} \, C(r;\mu) \!+\! S_{\varepsilon^{(\lambda)}}^{\,\dag} \, T(r;\mu) \right]  \vec{\tau}_{P}  \!\cdot\! \vec{\tau}_{N}, \nonumber \\
V_{P^{*}N \rightarrow P^{*}N} &\!=\!& \frac{g g_{\pi NN}}{\sqrt{2}m_{N}f_{\pi}} \frac{1}{3} \label{eq:pot_P*NP*N_ff} \\
&& \hspace{-0em} \times \left[ \vec{T}\!\cdot\!\vec{\sigma} \, C(r;m_{\pi}) \!+\! S_{T} \, T(r;m_{\pi}) \right] \vec{\tau}_{P}  \!\cdot\! \vec{\tau}_{N}, \nonumber
\end{eqnarray}
respectively.
Here $\vec{\tau}_{P}$ and $\vec{\tau}_{N}$ are isospin operators for $P^{(*)}$ and $N$. We define the operators 
$
S_{\varepsilon^{(\lambda)}} \!=\! 3 (\vec{\varepsilon}^{\,(\lambda)}\!\cdot\!\hat{r})(\vec{\sigma}\!\cdot\!\hat{r}) \!-\! \vec{\varepsilon}^{\,(\lambda)}\!\cdot\!\vec{\sigma}
$ and
$S_{T} \!=\! 3 (\vec{T}\!\cdot\!\hat{r})(\vec{\sigma}\!\cdot\!\hat{r}) \!-\! \vec{T}\!\cdot\!\vec{\sigma}$.
Note that the potential (\ref{eq:pot_PNP*N_ff}) includes the modified mass scale $\mu^{2}=m_{\pi}^{2}-(m_{P^{*}}-m_{P})^{2}$ because of the different masses $m_{P^{*}}$ and $m_{P}$ for $P^{*}$ and $P$.
In order to estimate the size effect of the nucleon and heavy meson, we introduce form factors, $(\Lambda_{N}^{2}-m_{\pi}^{2})/(\Lambda_{N}^{2}+\vec{q}^{\,\,2})$ and $(\Lambda_{P}^{2}-m_{\pi}^{2})/(\Lambda_{P}^{2}+\vec{q}^{\,\,2})$, in the momentum space at each vertex for $\pi NN$ and $\pi P^{(*)}P^{*}$, respectively.
$\vec{q}$ is the momentum of the propagating pion, and $m_{\pi}$ is the pion mass  (replaced to $\mu$ in Eq.~(\ref{eq:pot_PNP*N_ff})). $\Lambda_{N}$ and $\Lambda_{P}$ are the cut off parameters for the nucleon and heavy meson, respectively.
Then, $C(r;m)$ and $T(r;m)$ in Eqs.~(\ref{eq:pot_PNP*N_ff}) and (\ref{eq:pot_P*NP*N_ff}) are defined as
\begin{eqnarray}
\hspace{-3em}&&C(r;m) \!=\! \int \frac{\mbox{d}^{3}p}{(2\pi)^3} \frac{1}{\vec{q}^{\,\,2}+m^{2}} 
 e^{i\vec{q} \cdot \vec{r}} \, 
F(\vec{q};m), \\
\hspace{-3em}&&T(r;m) S_{12}(\hat{r}) \!=\! \int \frac{\mbox{d}^{3}p}{(2\pi)^3} \frac{- \vec{q}^{\,\,2}}{\vec{q}^{\,\,2}+m^{2}} 
S_{12}(\hat{q})e^{i\vec{q} \cdot \vec{r}} F(\vec{q};m),
\end{eqnarray}
with $S_{12}(\hat{x}) \!=\! 3 (\vec{\sigma}_{1} \cdot \hat{x}) (\vec{\sigma}_{2} \cdot \hat{x}) - \vec{\sigma}_{1} \cdot \vec{\sigma}_{2}$, and $F(\vec{q};m)\!=\! (\Lambda_{N}^{2} \!-\! m^{2})/(\Lambda_{N}^{2} \!+\! \vec{q}^{\,\,2}) \!\times\! (\Lambda_{P}^{2} \!-\! m^{2})/(\Lambda_{P}^{2} \!+\! \vec{q}^{\,\,2})$.

Now we move to discussion about the $P^{(*)}N$ bound states.
In this analysis, we consider four states, whose quantum numbers are classified to $J^{P}\!=\!1/2^{-}$ and $3/2^{-}$ with isospin $I\!=\!0$ and $1$.
The $1/2^{-}$ states with $I\!=\!0$ and 1 are superpositions of three states, $^{2}S_{1/2}$ for $PN$, and $^{2}S_{1/2}$ and $^{4}D_{1/2}$ for $P^{*}N$ with the standard notation $^{2S+1}L_{J}$.
The $3/2^{-}$ states with $I\!=\!0$ and 1 are superpositions of four states, $^{2}D_{3/2}$ for $PN$, and $^{4}S_{3/2}$, $^{4}D_{3/2}$, and $^{2}D_{3/2}$ for $P^{*}N$. 
A similar analysis has been done in Ref.~\cite{Cohen:2005bx}, in which they however do not take into account the mixing between different spin states, $^{2}S_{1/2}$ in $PN$ and $P^{*}N$, and $^{4}D_{1/2}$ in $P^{*}N$.


\begin{figure}
      \includegraphics[scale=0.4,clip]{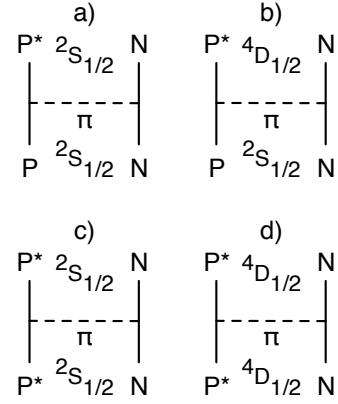} 
    \caption{\small \baselineskip=0.5cm The schematic diagrams of each component in the potential $V_{1/2^{-}}$ in Eq.~(\ref{eq:potential_1/2}). See the text for details.}
    \label{fig:potential}
\end{figure}

\begin{figure*}[t]
      \includegraphics[scale=0.85,clip]{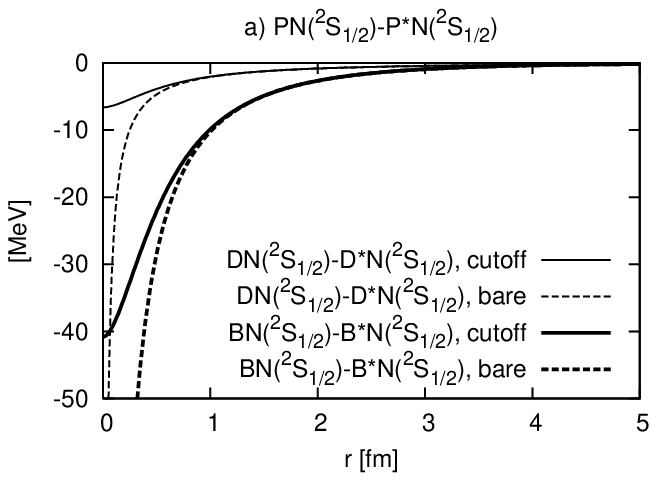} 
      \includegraphics[scale=0.85,clip]{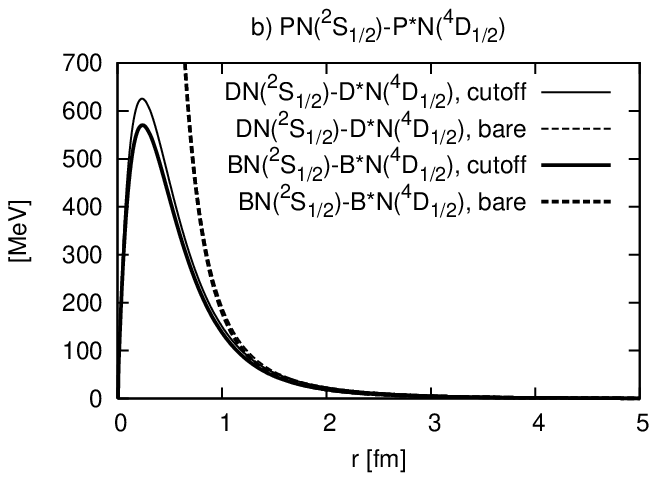} \\
      \includegraphics[scale=0.85,clip]{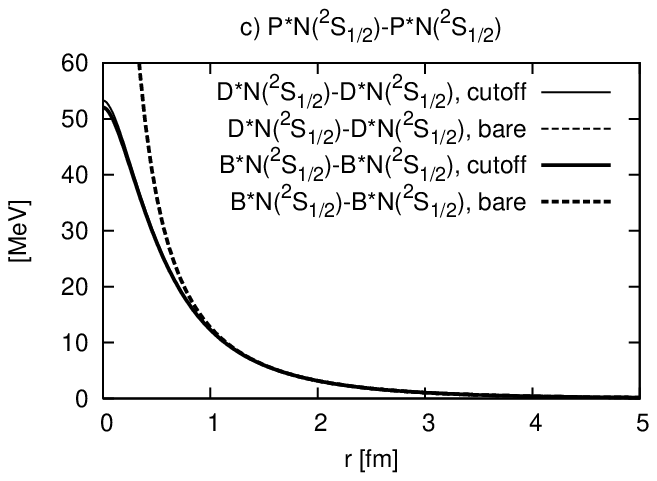} 
      \includegraphics[scale=0.85,clip]{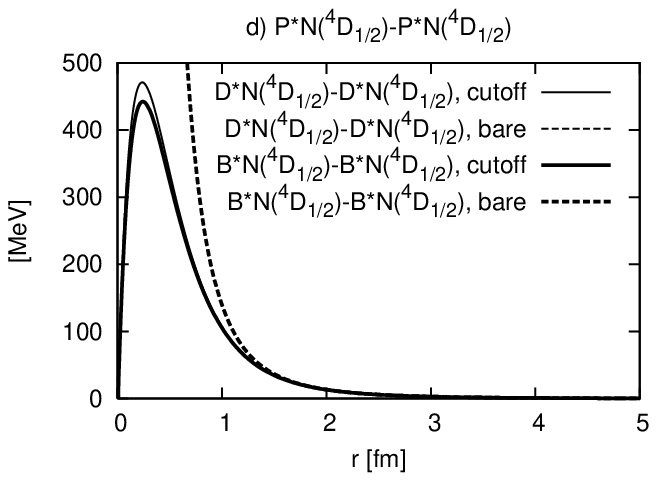} \\
    \caption{\small \baselineskip=0.5cm The plots of each component of the potential $V_{1/2^{-}}$ with $I\!=\!0$ in Eq.~(\ref{eq:potential_1/2}). See the text for details.}
    \label{fig:pot_half}
\end{figure*}

With these basis, we explicitly represent the potentials (\ref{eq:pot_PNP*N_ff}) and (\ref{eq:pot_P*NP*N_ff}) as
\begin{eqnarray}
V_{1/2^{-}} \!&=&\! \frac{g g_{\pi NN}}{\sqrt{2}m_{N}f_{\pi}} \frac{1}{3} \label{eq:potential_1/2} \\
&& \hspace{-1em}
 \times \left(
\begin{array}{ccc}
 0 & \sqrt{3}C_{\mu}  & -\sqrt{6}T_{\mu} \\
 \sqrt{3}C_{\mu} & -2C_{m_{\pi}} & 0 \\
  -\sqrt{6}T_{\mu} & 0  & C_{m_{\pi}}-2T_{m_{\pi}} 
\end{array}
\right)\vec{\tau}_{P}  \!\cdot\! \vec{\tau}_{N}, \nonumber \\
V_{3/2^{-}} \!&=&\! \frac{g g_{\pi NN}}{\sqrt{2}m_{N}f_{\pi}} \frac{1}{3} \label{eq:potential_3/2} \\
&& \hspace{-1em}
 \times \left(
\begin{array}{cccc}
0 & \sqrt{3}T_{\mu} & -\sqrt{3}T_{\mu} & \sqrt{3}C_{\mu} \\
 \sqrt{3}T_{\mu} & C_{m_{\pi}} & 2 T_{m_{\pi}} & T_{m_{\pi}}\\
 -\sqrt{3}T_{\mu} & 2 T_{m_{\pi}} & C_{m_{\pi}} & - T_{m_{\pi}} \\
 \sqrt{3}C_{\mu} & T_{m_{\pi}} & - T_{m_{\pi}} & -2 C_{m_{\pi}}
\end{array}
\right) \vec{\tau}_{P}  \!\cdot\! \vec{\tau}_{N}, \nonumber
\label{eq:pot_matrix}
\end{eqnarray}
for $J^{P} \!=\!1/2^{-}$ and $3/2^{-}$, respectively, with the abbreviation $C_{m}\!=\!C(r;m)$ and $T_{m}\!=\!T(r;m)$ for $m\!=\!m_{\pi}$ and $\mu$. 
We confirm that the mixing among the different spin and angular momentum states is given in the off-diagonal components in the matrices given above.
Concerning the independnent components in $V_{1/2^{-}}$, 
schematic diagrams are shown in Fig.~\ref{fig:potential}; a) $PN(^{2}S_{1/2})$-$P^{*}N(^{2}S_{1/2})$, b) $PN(^{2}S_{1/2})$-$P^{*}N(^{4}D_{1/2})$, c) $PN(^{2}S_{1/2})$-$PN(^{2}S_{1/2})$, and d) $P^{*}N(^{4}D_{1/2})$-$P^{*}N(^{4}D_{1/2})$.
The kinetic terms are
\begin{eqnarray}
K_{1/2^{-}}\!&=&\! \mbox{diag}\left(
 -\frac{1}{2\widetilde{m}_{P}} \triangle_{0}, \right.
-\frac{1}{2\widetilde{m}_{P^*}} \triangle_{0} + \Delta m_{PP^{*}}, \\
&&\left. -\frac{1}{2\widetilde{m}_{P^*}} \triangle_{2} + \Delta m_{PP^{*}} 
\right), \nonumber \\
K_{3/2^{-}}\!&=&\! \mbox{diag}\left(
 -\frac{1}{2\widetilde{m}_{P}} \triangle_{2}, \right.
-\frac{1}{2\widetilde{m}_{P^*}} \triangle_{0} + \Delta m_{PP^{*}}, \\
 && -\frac{1}{2\widetilde{m}_{P^*}} \triangle_{2}  + \Delta m_{PP^{*}},
\left. -\frac{1}{2\widetilde{m}_{P^*}} \triangle_{2} + \Delta m_{PP^{*}} \nonumber
\right),
\end{eqnarray}
for $J^{P} \!=\!1/2^{-}$ and $3/2^{-}$, respectively.
Here we define $\triangle_{0}=\partial^{2}/\partial r^{2} + (2/r) \partial/\partial r$ and $\triangle_{2}=\triangle_{0}+ 6/r^2$, $\widetilde{m}_{P^{(*)}} \!=\! m_{N}\,m_{P^{(*)}}/(m_{N} \!+\! m_{P^{(*)}})$, and $\Delta m_{PP^{*}}\!=\!m_{P^{*}}\!-\!m_{P}$.
The eigenvalue equation with the given hamiltonian, $H_{J^{P}}=K_{J^{P}}+V_{J^{P}}$ with $J^{P}=1/2^{-}$ and $3/2^{-}$, is numerically solved by a variational method.
The binding energy is realized as a difference from the threshold $m_{N}+m_{P}$.

We fix the cut off parameters $\Lambda_{N}$ and $\Lambda_{P}$, and the sign of the coupling constant $g$.
According to Eq.~(\ref{eq:piNN}), the OPEP between two nucleons is obtained by
\begin{eqnarray}
V_{NN} \!=\! \left( \frac{g_{\pi NN}}{2m_{N}} \right)^2 \frac{1}{3}
\left[ \vec{\sigma}_1\!\cdot\!\vec{\sigma}_2 C_{m_{\pi}} \!+\! S_{12}(\hat{r}) T_{m_{\pi}})\right] \vec{\tau}_{1} \!\cdot\! \vec{\tau}_{2}.
\end{eqnarray}
We find that $\Lambda_{N}\!=\!940$ MeV reproduces a deuteron state with the binding energy 2.2 MeV and the relative radius 3.7 fm.
For $\Lambda_{P}$, we assume the relation, $\Lambda_{P}/\Lambda_{N}\!=\!r_{N}/r_{P}$, in terms of matter radii of the nucleon $r_{N}$ and heavy meson $r_{P}$, because the cut off parameter represents the inverse of the size of hadrons.
In the constituent quark model, the sizes $r_{N}$ and $r_{P}$ are characterized by the frequencies $\omega_N$ for nucleon and $\omega_{P}$ for heavy meson, respectively, in harmonic oscillator potentials \cite{Oh:2009zj}.
In order to evaluate $\omega_{N}$ and $\omega_{P}$, we make use of the charge radii which are relatively well known in experiments.
Thus, $\omega_{N}$ is determined to reproduce the charge
 radius with subtraction of the pion cloud, which is estimated to be around 2/3 of the observed value $0.875$ fm \cite{Amsler:2008zzb}, namely $0.58$ fm, in the analysis in the chiral quark model \cite{Oset:1984tva}.
Concerning the charge radii of heavy mesons, we use $0.43$ fm for $D^{+}$ and $0.62$ fm  for $B^{+}$ \cite{Hwang:2001th}.
With the constituent quark masses $m_{u,d} \!=\! 300$ MeV, $m_{c} \!=\! 1500$ MeV, and $m_{b} \!=\! 4700$ MeV, we obtain 
$\omega_{N} \!=\! 420$ MeV and $\omega_{P} \!=\! 330$ MeV.
Finally, we obtain the ratios $r_{N}/r_{D} \!=\! 1.35$ and $r_{N}/r_{B} \!=\! 1.29$, and then $\Lambda_{D}\!=\!1266$ MeV and $\Lambda_{B}\!=\!1213$ MeV.
As for the sign of $g$, we use $g\!=\!-0.59$.
If the opposite sign is used, the binding energy becomes slightly larger by a few MeV.
From the point of view of G-parity, such states would be assigned to $DN$ and $\bar{B}N$.

With these parameters, we obtain the potentials $V_{1/2^{-}}$ and $V_{3/2^{-}}$ in Eqs.~(\ref{eq:potential_1/2}) and (\ref{eq:potential_3/2}).
For example, we plot each component in the potential $V_{1/2^{-}}$ with $I \!=\! 0$ in Fig.~\ref{fig:pot_half}.
Each figure corresponds to the diagrams in Fig.~\ref{fig:potential}.
The narrow (thick) curves show the $\bar{D}^{(*)}N$ ($B^{(*)}N$) potentials, and the solid (dashed) curves show the cases with (without) the form factor.


\begin{figure}
      \includegraphics[scale=1.0,clip]{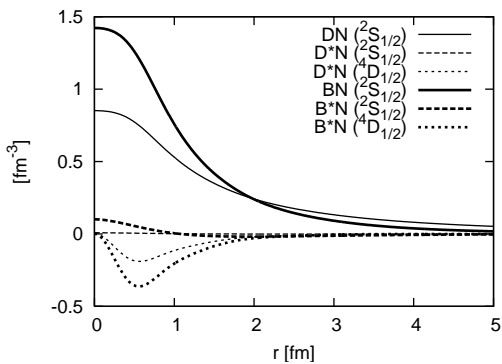} 
    \caption{\small \baselineskip=0.5cm The wave functions of the $\bar{D}N$ and $BN$ bound states for $J^{P} \!=\! 1/2^-$ with $I \!=\! 0$.}
    \label{fig:wf_DN}
\end{figure}

By using these potentials, we find the $\bar{D}N$ and $BN$ bound state solutions for $J^{P} \!=\! 1/2^-$ with $I \!=\! 0$. 
Their binding energies and the relative radii are 1.4 MeV and 9.4 MeV, and 3.8 fm and 1.7 fm, respectively, which are summarized in Table.~\ref{table:binding_energy}. 
The wave functions are shown in Fig.~\ref{fig:wf_DN}, where the solid, dashed, and dotted curves represent $PN(^2 S_{1/2})$, $P^{*}N(^2 S_{1/2})$, and $P^{*}N(^4 D_{1/2})$, respectively.
The narrow (thick) curves show the $\bar{D}N$ ($BN$) states.
We indicate that the $P^{*}N(^{4}D_{1/2})$ component is crucially important, though it is smaller than the $PN(^{2}S_{1/2})$ as shown in Fig.~\ref{fig:wf_DN}.
The binding energy is induced mainly by the tensor coupling between $PN(^2 S_{1/2})$ and $P^{*}N(^4 D_{1/2})$, which are in off-diagonal terms in $V_{1/2^{-}}$ in Eq.~(\ref{eq:potential_1/2}).
Indeed, without this tensor coupling, we cannot obtain any bound state.
For more details, one can see the $PN(^{2}S_{1/2})$-$PN(^{2}S_{1/2})$ and $P^{*}N(^{4}D_{1/2})$-$P^{*}N(^{4}D_{1/2})$ components which are diagonal terms in the potential $V_{1/2^{-}}$ are repulsive as shown in Fig.~\ref{fig:pot_half}c) and d).
Obviously, the bound states are not realized only by these terms.
However, the $PN(^{2}S_{1/2})$-$P^{*}N(^{4}D_{1/2})$ component with the tensor coupling which is an off-diagonal term in $V_{1/2^{-}}$ is stronger than the diagonal ones.
As a result, the off-diagonal component, causing the mixing of the $PN(^{2}S_{1/2})$ and $P^{*}N(^{4}D_{1/2})$ states, induces the attraction.

Thus, the $PN$-$P^{*}N$ mixing plays an essential role in these systems.
We see that the $BN$ state is more deeply bound and compact than the $\bar{D}N$ state.
This is because the the smaller mass splitting between $B$ and $B^{*}$ strengthens the $BN$-$B^{*}N$ mixing, and the kinetic energy of $B$ mesons is smaller.
In the heavy quark limit, the complete mass degeneracy of $P$ and $P^{*}$ induces the ideal $PN$-$P^{*}N$ mixing, which would give the maximal binding energy.

It is worth to emphasize that we can not find $\bar{D}N$ and $BN$ bound states for other channels, $J^{P} \!=\! 1/2^-$ with $I\!=\!1$, and $J^{P} \!=\! 3/2^{-}$ with $I\!=\!0$ or 1.
As a result, we conclude that $J^P \!=\! 1/2^-$ with $I \!=\! 0$ are the most promising channel for detecting stable $\bar{D}N$ and $BN$ bound states. 

In experiments, the $\bar{D}N$ and $BN$ bound states would be searched in $e^{+}e^{-}$ collisions, or anti-proton beam with deuteron targets \cite{Sibirtsev:2000yy,Haidenbauer:2008ff}.
Because they do not decay in strong interaction, they are experimentally well accessible despite of their small binding energies.
The $\bar{D}N$ and $BN$ wave functions have the following components,
\begin{eqnarray}
\hspace{-1.5em}|\bar{D}N \rangle \!&=&\!
c^{D}_{0} \left( | D^{-}p \rangle \!-\! | \bar{D}^{0}n \rangle \right) \!+\!
c^{D}_{1} \left( | D^{*-}p \rangle \!-\! | \bar{D}^{*0}n \rangle \right),
 \\
\hspace{-1.5em}|BN \rangle \!&=&\!
c^{B}_{0} \left( | B^{0}p \rangle \!-\! | B^{+}n \rangle \right) \!+\!
c^{B}_{1} \left( | B^{*0}p \rangle \!-\! | B^{*+}n \rangle \right),
\end{eqnarray}
with some coefficients $c_{i}^{H}$ ($i=0, 1$ and $H=D, B$).
The weak decay processes, $D^{-}p \rightarrow K^{+}\pi^{-} \pi^{-} + p$ and $B^{0}p \rightarrow D^{-}\pi^{+} + p$, in the first component would be available for reconstruction of the invariant mass.

\begin{table}[t]
\caption{\small \baselineskip=0.5cm The properties of the $\bar{D}N$ and $BN$ bound states for $J^{P}\!=\!1/2^{-}$ with $I\!=\!0$.}
\begin{center}
\begin{tabular}{c@{\hspace*{0.5cm}}c@{\hspace*{0.5cm}}c}
\hline
 & $\bar{D}N$ & $BN$ \\
\hline
binding energy & 1.4 MeV & 9.4 MeV \\
\hline
relative radius & 3.8 fm & 1.7 fm \\
\hline
\end{tabular}
\end{center}
\label{table:binding_energy}
\end{table}%

We mention that, contrary to the $\bar{D}(B)N$ states, the $D(\bar{B})N$ states is more complicated.
Although the $u$-channel would be suppressed due to a heavy $Q\bar{Q}$ pair creation, the $s$-channel is not necessarily small.
Concerning the Born term, it is phenomenologically known in $\bar{K}N$ scatterings that the $s$-channel plays a minor role \cite{MuellerGroeling:1990cw}.
However, the $\Lambda_{c(b)}^{*}$ and $\pi \, \Sigma_{c(b)}$ states below thresholds may strongly couple to the $D(\bar{B})N$ states.
Therefore, the coupled channel effects should be included. 
We note that the $\bar{D}(B)N$ states discussed here are not affected by the presence of $\Lambda_{c(b)}^{*}$ nor $\pi \, \Sigma_{c(b)}$.

As further theoretical studies, in addition to OPEP, we may include chiral loops, scalar and vector meson exchanges, and direct quark exchanges.
However, we consider that the ranges of their interactions would be smaller than the sizes of the $\bar{D}N$ and $BN$ states, hence they would be minor.
Nevertheless, the vector meson exchange is interesting because it is compatible with the Weinberg-Tomozawa type interaction in chiral symmetry \cite{Sudoh}.
More information about the interactions will be provided also from the analysis of molecular picture of the recently observed exotic charmed mesons \cite{AlFiky:2005jd}, such as X(3872) \cite{Choi:2003ue}, Z$^{\pm}$(4430) \cite{:2007wga}, and so forth.

The stability of $\bar{D}N$ and $BN$ states should be compared with the instability of $KN$ states as candidate of pentaquarks.
It has been theoretically studied that the $K$ meson is not bound with a nucleon by the Weinberg-Tomozawa interaction with chiral symmetry \cite{Hyodo:2006yk}.
In contrast, our analysis shows the $\bar{D}$ and $B$ mesons are bound with a nucleon by OPEP with respecting heavy quark symmetry.
Therefore, the $\bar{D}N$ and $BN$ bound states may open another way to search for pentaquark with heavy flavors \cite{Oh:1994yv}.

The work on $\bar{D}N$ bound states is extended to exotic nuclei containing $\bar{D}$ mesons.
One can expect that the binding energy of the $\bar{D}$ meson in such exotic nuclei becomes larger as baryon number increases.
Such exotic nuclei will be investigated in J-PARC and GSI by using anti-proton beams with targets of nuclei \cite{Golubeva:2002au,Sibirtsev:1999js}.
Let us discuss the $\bar{D}NN$ states with baryon number two.
The possible states for $I\!=\!1/2$ are classified to $| \bar{D}(NN)_{0,1} \rangle \!+\! | \bar{D}^{*}(NN)_{1,0} \rangle$ with $J^{P}\!=\!0^{-}$ and $| \bar{D}(NN)_{1,0} \rangle + | \bar{D}^{*}(NN)_{1,0} \rangle + | \bar{D}^{*}(NN)_{0,1} \rangle$ with $J^{P}\!=\!1^{-}$. 
For $I\!=\!3/2$, they are $|\bar{D}(NN)_{0,1} \rangle$ with $J^{P}\!=\!0^{-}$ and $|\bar{D}^{*}(NN)_{0,1} \rangle$ with $J^{P}\!=\!1^{-}$,  where subscripts denote the spin and isospin of nucleon pairs.
Only the former two states for $I\!=\!1/2$ contain the attractive $\bar{D}N$ pairs for $J^{P}\!=\!1/2^{-}$ with $I\!=\!0$.
Therefore, one can expect that the $\bar{D}NN$ states, as well as the $BNN$ states, for $J^{P}\!=\!0^{-}$ and/or $1^{-}$ with $I\!=\!1/2$ would be stable.

It is valuable to compare the $\bar{D}NN$ and $BNN$ states with the $K^{-}pp$ states ($\bar{K}(NN)_{0,1}$ in our notation) \cite{Dote:2008hw}.
The $K^{-}pp$ state is considered to have large binding energy 20-70 MeV.
However, the $K^{-}pp$ state decays to $\pi Y N$ by a strong interaction with the decay width 40-70 MeV.
On the other hand, the $\bar{D}NN$ and $BNN$ states are stable in the strong decay as discussed above.
Therefore, the $\bar{D}NN$ and $BNN$ states may provide more precise information about exotic nuclei.
More quantitative analyses including few-body calculation will be presented in Ref. \cite{Sudoh}.

The charmed and bottom nuclei like $\bar{D}NN$ and $BNN$ states are comparable with the hypothetical existence of the kaonic nuclei, in which $K$ mesons are bound in nuclei \cite{Goldhaber}.
The existence of such nuclei may realize if there is a sufficient number of nucleons around $K$ meson, though $KN$ interaction itself seems not attractive to form $KN$ molecule \cite{Hyodo:2006yk}.
We stress again that in our analysis the $\bar{D}$ and $B$ mesons have sufficiently large attraction to form the stable charmed and bottom nuclei beginning with baryon number one.

In summary, bound states of nucleon and an open heavy flavor meson are discussed with respecting the heavy quark symmetry.
It is found $\bar{D}N$ and $BN$ bound states with binding energies 1.4 MeV and 9.4 MeV, respectively, for $J^{P} \!=\! 1/2^{-}$ with $I \!=\! 0$, and no bound states in other channels.
These states are stable in the strong decay, and can be observed in the weak decay processes $\bar{D} N \rightarrow K^{+}\pi^{-} \pi^{-} + p$, and $B N \rightarrow D^{-}\pi^{+} + p$.
The existence of $\bar{D}N$ and $BN$ bound states would provide an opportunity to probe new exotic states near the thresholds, and open a new way to investigate for exotic nuclei with variety of multi-flavor explored at future hadron facilities such as J-PARC and GSI.

The authors thank A. Dote for fruitful discussions.


\end{document}